# The Simons Observatory: Development and Optical Evaluation of Achromatic Half-Wave Plates

**Junna Sugiyama**[1] · **Tomoki Terasaki**[1] · **Kana Sakaguri**[1] · **Bryce Bixler**[2] · **Yuki Sakurai**[3,4] · **Kam Arnold**[2] · **Kevin T. Crowley**[2] · **Rahul Datta**[5] · **Nicholas Galitzki**[6,7] · **Masaya Hasegawa**[8] · **Bradley R. Johnson**[9] · **Brian Keating**[2] · **Akito Kusaka**[1,4,10] · **Adrian Lee**[10,11] · **Tomotake Matsumura**[4] · **Jeffrey Mcmahon**[5] · **Maximiliano Silva-Feaver**[2] · **Yuhan Wang**[12] · **Kyohei Yamada**[1]

## Abstract

The Simons Observatory (SO) experiment is a cosmic microwave background (CMB) experiment located in the Atacama Desert, Chile. The SO's small aperture telescopes (SATs) consist of three telescopes designed for precise CMB polarimetry at large angular scales. Each SAT uses a cryogenic rotating half-wave plate (HWP) as a polarization modulator to mitigate atmospheric $1/f$ noise and other systematics. To realize efficient polarization modulation over the observation bands, we fabricated an achromatic HWP (AHWP) consisting of three sapphire plates with anti-reflection coatings. The AHWP is designed to have broadband modulation efficiency and transmittance. This paper reports on the design and the preliminary characterization of the AHWPs for SATs.

## 1 Introduction

The cosmic microwave background (CMB) polarization has rich information of the early universe. One of the largest noise sources for the ground-based CMB experiments is the unpolarized low-frequency noise, also known as $1/f$ noise, which is caused by atmospheric fluctuations. To improve polarization sensitivity, it is effective to modulate the polarized signals to frequencies above the knee frequency (∼ 1 Hz) of the $1/f$ noise. Several CMB experiments, especially the ground-based ones, have used polarization modulators in the past [1–12]. The Simons Observatory (SO) [13] will also introduce the polarization modulators to its telescopes called small aperture telescopes (SATs) [14]. The SATs consist of three telescopes, two

Extended author information available on the last page of the article

Springer

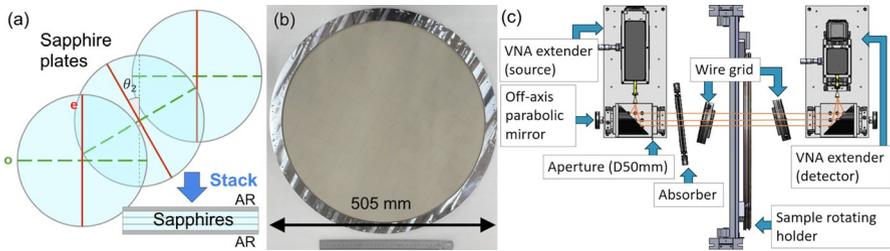

**Fig. 1 a** The sapphire HWP stack model. The green/red lines show the ordinary/extra-ordinary axis of different indices ($n_o = 3.05 \pm 0.03$/$n_e = 3.38 \pm 0.03$) [10]. They are stacked into one AHWP with AR-coated alumina plates. **b** The AHWP view. The photo shows the AR-coated surface and the aluminum holder. **c** The configuration of the optical measurement. This setup is described in Ref. [26] in detail of which are dedicated to the 90/150 GHz bands (mid-frequency bands denoted as MF) and the remaining one to the 220/280 GHz bands (ultra high-frequency bands denoted UHF). The SATs use a continuously rotating half-wave plate (HWP) as a polarization modulator, which comprises the HWP optic and the cryogenic rotation mechanism. This paper describes the optical properties of HWP, while the cryogenic rotation mechanism is described elsewhere [15].

The modulated signal $d_m(t)$ is expressed in the HWP angular rotation speed $\omega_{\text{HWP}}$:

$$d_m(t) = T \cdot I(t) + \epsilon T \cdot \text{Re}[(Q(t) + iU(t))] \exp[-4i(\omega_{\text{HWP}} t + \phi_v)] + N(t); \quad (1)$$

where $t$ is time; $T$, $\epsilon$ and $\phi_v$ are the transmittance, polarization modulation efficiency and optical axis phase of the HWP; $I(t)$, $Q(t)$ and $U(t)$ are the Stokes parameters of the incident light and $N(t)$ is the noise on the detector, respectively.

Various designs of HWPs have been developed for CMB polarimetry [16–19]. The SATs use a three-layer achromatic HWP (AHWP), so-called Pancharatnam AHWP [20], to achieve broadband modulation efficiency. The surface of the AHWP is anti-reflection (AR) coated to improve the transmission. We report on the fabrication and preliminary evaluation of AHWPs for three SATs: Two MF SATs named SAT MF-1 and SAT MF-2, and SAT UHF.

## 2 Fabrication

We follow the same AHWP fabrication process as POLARBEAR-2b [21, 22]. The AHWP consists of three sapphire plates and two AR-coated alumina plates. Its optical diameter is 490 mm, the largest in CMB experiments to our knowledge. The sapphire plates are manufactured at Guizhou Haotian Optoelectronics Technology Co., Ltd. [23]. Sapphires for SAT UHF are ground to be thin and flat at KYOCERA [24]. The thickness of each sapphire plate is $3.75 \pm 0.01$ mm for MF-1 and MF-2; and $1.60 \pm 0.01$ mm for UHF. Three sapphire plates are stacked in the configuration described in Fig. 1a. The orientation of the center sapphire is shifted by $\theta_2$ (54 deg for MF-1 and MF-2; 57 deg for UHF) to optimize the broadband modulation efficiency for each SAT. The sapphire stack is placed between the AR-processed aluminas, and all the layers are glued at the center with Epo-Tek 301-2 [25]. The AHWP edge is held in place with an aluminum holder as shown in Fig. 1b. The MF-1 and UHF AHWPs use the two-layer AR coating described in Sakaguri et al. [26]. The MF-2 AHWP uses the metamaterial AR described in Golec et al. [27].

## 3 Optical Test at Room Temperature

The transmission and the modulation efficiency of the AHWPs are optically measured at room temperature. The optical axis phase $\phi_\nu$ in Eq. (1) depends on the microwave frequency $\nu$ for the Pancharatnam AHWP. We also measured this frequency-dependent phase $\phi_\nu$.

### 3.1 Setup

The AHWP response to linear polarization is evaluated using the setup shown in Fig. 1c. The vector network analyzer (VNA) measures the intensity and phase of the input/output signal. The angle of two wire grids is fixed vertically; one wire grid horizontally polarizes the incident electric field on the AHWP $\boldsymbol{E}_i = (E_i^h \neq 0, E_i^v = 0)$, and the other extracts $E_t^h$ from the transmitted electric field of $\boldsymbol{E}_t = (E_t^h, E_t^v)$. The AHWP is held on the rotating sample holder and is measured every 5 deg of the rotation angle to estimate the modulation efficiency and the phase shift. The planes of wire grids and the absorber are not aligned parallel with the AHWP to prevent incident and reflected light from making standing waves.

### 3.2 Results

Figure 2 shows the result for MF-1 AHWP. The measured modulation efficiency $\epsilon$ and the phase shift $\phi_\nu$ are shown in Fig. 2a. The AHWP has nearly 100% modulation efficiency over the observation bands with its optical axis well characterized. The measured and simulated room temperature transmission from $E_i^h$ to $E_t^h$ are compared in Fig. 2b and show good agreement. The simulations employ the method described in Ref. [28] and adopt the loss tangent values in the literature [10, 26, 29–33]. The transmittance is higher at lower temperatures because the absorption is lower. Expected transmission at 100 K is calculated using the AHWP loss tangent at 100 K with other parameters the same as 300 K. The cryogenic MF-1 AHWP has a transmission of > 90% over the observation bands. The fine fringes in the transmission spectra arise from the internal reflections in the birefringent sapphires and the small AR imperfection. The band-averaged modulation efficiency $\epsilon$, transmittance $T$ and reflectance $R$ for all AHWPs are shown in Table 1. The leftover of $1 - T - R$ stands for the absorption. As shown in Table 1, high band-averaged modulation efficiency and transmittance, and low reflectance are achieved. The operation temperature will be lower than 100 K, and the transmittance will slightly increase compared to these numbers.

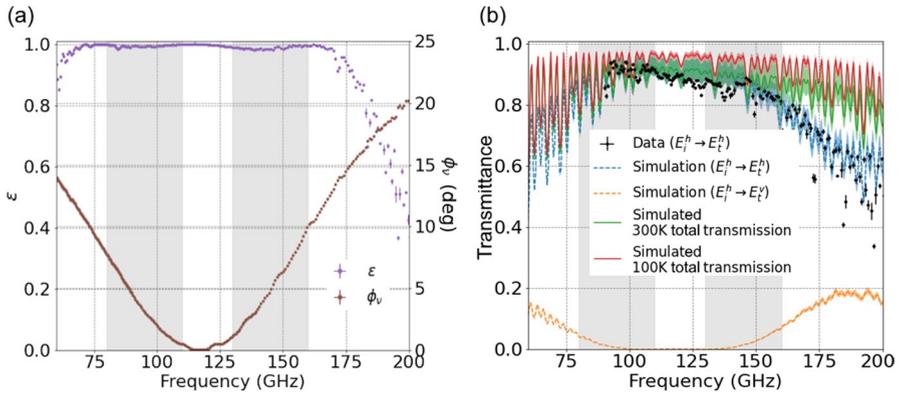

**Fig. 2** The measured MF-1 AHWP spectra. The gray bands show the observation frequencies. **a** Purple dots show the modulation efficiency $\epsilon$, and brown dots show the phase shift $\phi_v$ of the optical axis. **b** Data points are the measured $E_i^h$ to $E_t^h$ transmission, and dotted lines show the simulated $E_i^h$ to $E_t^h$ or $E_t^v$ transmission, and solid lines show the simulated 300 K or 100 K total transmission. These transmission spectra change with AHWP rotation

**Table 1** The modulation efficiency $\epsilon$, transmittance $T$ and reflectance $R$ of the SAT AHWPs

| Frequency band | MF-1 | | MF-2 | | UHF | |
| --- | --- | --- | --- | --- | --- | --- |
| | 90 GHz | 150 GHz | 90 GHz | 150 GHz | 220 GHz | 330 GHz |
| $\epsilon$ (%) | 99.5 ± 0.03 | 98.8 ± 0.02 | 97.3 ± 0.01 | 99.0 ± 0.01 | 99.8 ± 0.02 | 99.7 ± 0.03 |
| $T$ at 300 K | 0.89 ± 0.04 | 0.84 ± 0.02 | 0.91 ± 0.03 | 0.84 ± 0.06 | 0.73 ± 0.02 | 0.72 ± 0.02 |
| $T$ at 100 K | 0.91 ± 0.02 | 0.92 ± 0.02 | 0.97 ± 0.01 | 0.91 ± 0.02 | 0.88 ± 0.01 | 0.87 ± 0.01 |
| $R$ | 0.07 ± 0.01 | 0.02 ± 0.01 | 0.02 ± 0.01 | 0.09 ± 0.02 | 0.03 ± 0.01 | 0.02 ± 0.01 |

$\epsilon$ is the optically measured value, while $T$ and $R$ are the simulated values. The leftover of $1 - T - R$ stands for the absorption. $T$ at 100 K is higher than at 300 K since the absorption decreases

## 4 Instrumental Polarization

### 4.1 Setup

We conducted an instrumental polarization (IP) measurement, a spurious polarization generated in a telescope [34, 35]. Here, we focus on IP moduled with frequency of four times HWP speed, which cause leakage into the polarized signal of interest. The left panel of Fig. 3 shows overall setup of the IP measurement conducted in the final integration test of the first SAT, SAT-MF1 [36]. We put a room temperature blackbody right on top of the window of SAT. Each detector sees the blackbody with an incidence angle that is proportional to its displacement from the center of the focalplane. The optical configuration inside of SAT-MF1 is the same as the on-site configuration except for neutral density (ND) filter (Eccosorb MF-112 [37]) in front of the detector modules, which attenuates in band emission from 300 K black body and prevents detectors from saturating. The radiation from the blackbody goes through the vacuum window, metal-mesh infrared (IR) filter, 40 K alumina IR filter, AHWP, 4 K alumina IR filter, optics tube, which mechanically supports three silicon lenses and a low-pass edge filter, another low-pass edge filter in front of detectors, and ND filter, and is detected by transition edge detectors (TESes) on focalplane [38]. The IP is observed by each TES as the Q and U of Eq. (1) and is recovered with a standard demodulation process [9, 35].

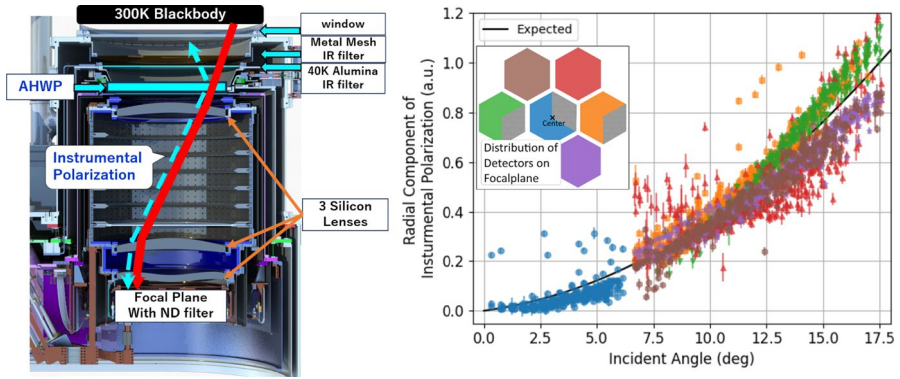

**Fig. 3** (Left) The configuration of the IP measurement. (Right) Each data point corresponds to amplitude of the radial component of IP. Color of focalplane corresponds to the color of each data point. Solid black line shows normalized quadratic behavior

### 4.2 Results

The observed small IP can be expressed as a sum of uniform polarization and radial polarization pattern across the focalplane. The uniform polarization component is not expected to be due to the AHWP because IP from AHWP has a radial property [34].[1] The right panel of Fig. 3 shows the radial component for 150 GHz as a function of incidence angle. Measured data were relatively calibrated among detectors by comparing measured saturation power of TESes with warm (∼ 300 K) load and cold (∼ 77 K) load on top of the window. The measured radial IP dependence shows reasonable agreement with expected quadratic dependence (shown in a black line) due to the main cause of IP, the differential transmission between s-wave and p-wave [34] at optical components such as AHWP, 40 K alumina filter and window. It means that AHWP is well fabricated and does not have any obvious defects. Difference among data points, orange/green points and red/purple points, could be explained by the calibration error due to that cold load being located 450 mm above the window whereas warm load was just on top of the window. We emphasize that this measurement was conducted about 1 year before the deployment of the telescope and improved our confidence on the AHWP. More precise IP characterization will be executed on-sky.

---

[1] While an imperfection of AHWP can make a non-radial IP, such as AR-coating non-uniformity, this type of IP should be observed with orthogonal polarization angles by orthogonal pair detectors. This is contrary our observation.

## 5 Summary

We evaluated the performance of the AHWPs from the warm and cold tests. In the optical measurements at room temperature, we characterized the transmittance, modulation efficiency and phase shift of the optical axis of the three AHWPs for SAT MF-1, MF-2 (90/150 GHz) and UHF (220/280 GHz). All the AHWPs achieved modulation efficiency of > 97%. The MF-1 and MF-2 AHWPs will have transmittance > 91% at < 100 K. The UHF AHWP will have transmittance > 87% at < 100 K. We also measured instrumental polarization in an integration test and confirmed the expected radial dependence, demonstrating that AHWP does not have any obvious defect.

**Acknowledgements** This work was supported in part by the Simons Foundation (Award #457687, B.K.). This work was also supported by JSR Fellowship, the University of Tokyo; and by FoPM and IGPEES, WINGS Program, the University of Tokyo. This work was also supported by JSPS Core-to-Core program Grant No. JPJSCCA20200003 and JSPS KAKENHI Grant Nos. JP23KJ0501, JP23H01202, 18H05539, 19H00674 and 23H00105 and International Research Center Formation Program to Accelerate Okayama University Reform (RECTOR).

**Funding** Open Access funding provided by The University of Tokyo.

## Declarations

**Conflict of interest** The authors declare no competing interests.

## Authors and Affiliations

**Junna Sugiyama[1] · Tomoki Terasaki[1] · Kana Sakaguri[1] · Bryce Bixler[2] · Yuki Sakurai[3,4] · Kam Arnold[2] · Kevin T. Crowley[2] · Rahul Datta[5] · Nicholas Galitzki[6,7] · Masaya Hasegawa[8] · Bradley R. Johnson[9] · Brian Keating[2] · Akito Kusaka[1,4,10] · Adrian Lee[10,11] · Tomotake Matsumura[4] · Jeffrey Mcmahon[5] · Maximiliano Silva-Feaver[2] · Yuhan Wang[12] · Kyohei Yamada[1]**

✉ Junna Sugiyama
  junna.sugiyama@phys.s.u-tokyo.ac.jp

✉ Tomoki Terasaki
  tomoki.terasaki@phys.s.u-tokyo.ac.jp

1. Department of Physics, The University of Tokyo, 7-3-1, Hongo, Bunkyo-ku, Tokyo 113-0033, Japan

2. Department of Physics, University of California, La Jolla, San Diego, CA 92093, USA

3. Graduate School of Natural Science and Technology, Okayama University, 3-1-1, Tsusima-Naka, Kita-ku, Okayama 700-8530, Japan

4. Kavli IPMU, The University of Tokyo, 5-1-5, Kashiwanoha, Kashiwa, Chiba 277-8583, Japan

5. Department of Astronomy and Astrophysics, University of Chicago, 5640 South Ellis Avenue, Chicago, IL 60637, USA

6. Department of Physics, University of Texas at Austin, Austin, TX 78712, USA

7. Weinberg Institute for Theoretical Physics, Texas Center for Cosmology and Astroparticle Physics, Austin, TX 78712, USA

8. High Energy Accelerator Research Organization, 1-1, Ooho, Tsukuba, Ibaraki 305-0801, Japan

9. Department of Astronomy, University of Virginia, Charlottesville, VA 22904, USA

10. Physics Division, Lawrence Berkeley National Laboratory, Berkeley, CA 94720, USA

11. Department of Physics, University of California, Berkeley, CA 94720, USA

12. Joseph Henry Laboratories of Physics, Princeton University, Princeton, NJ 08544, USA